\documentclass[aps,prl,reprint,superscriptaddress]{revtex4-1}
\usepackage{blindtext}
\usepackage{xcolor}
\usepackage{hyperref}
\usepackage{epsfig}
\usepackage{subfigure}
\usepackage{graphicx}
\usepackage[toc,page]{appendix}
\usepackage{epstopdf}
\usepackage{amsmath,amssymb,amsfonts}
\usepackage{array}
\usepackage{url}
\usepackage{hyperref}
\usepackage{lineno}
\usepackage{xspace}
\usepackage{listings}
\usepackage{float}
\usepackage{subfigure}
\usepackage{subfiles}

\graphicspath{ {Figures_list/} }

\begin{document}
\title{Precision Measurement of (Net-)proton Number Fluctuations in Au+Au Collisions at RHIC}
\author{(The STAR Collaboration)}
\date{\today}

\begin{abstract}

We report precision measurements on cumulants ($C_{n}$) and factorial cumulants ($\kappa_{n}$) of (net-)proton number distributions up to fourth-order in Au+Au collisions from phase II of the Beam Energy Scan program at RHIC. (Anti-)protons are selected at midrapidity ($|y|<0.5$) within a transverse momentum range of $0.4 < p_T < 2.0$ GeV/$c$. The collision energy and centrality dependence of these cumulants are studied over center-of-mass energies $\sqrt{s_{NN}}$ = 7.7 -- 27 GeV. Relative to various non-critical-point model calculations and peripheral collision 70-80\% data, the net-proton $C_4/C_2$ measurement in 0-5\% collisions shows a minimum around 19.6 GeV for significance of deviation at $\sim2$ -- $5\sigma$.
In addition, deviations from non-critical baselines around the same collision energy region are also seen in proton factorial cumulant ratios, especially in $\kappa_2/\kappa_1$ and $\kappa_3/\kappa_1$. Dynamical model calculations including a critical point are called for in order to understand these precision measurements. 

\end{abstract}

\maketitle


A key focus of experiments involving the collision of heavy ions at high energy has long been the search for a possible critical point associated with the phase transitions of matter interacting via the strong force. In the phase diagram of such matter, known as the quantum chromodynamics (QCD) phase diagram, the critical point marks the boundary between a crossover quark-hadron transition at low baryochemical potential ($\mu_B$) and a possible first-order transition at higher $\mu_B$~\cite{Rajagopal:2000wf,Bzdak:2019pkr,Pandav:2022xxx}. While first-principle calculations based on lattice QCD have established the crossover at small $\mu_B$~\cite{Aoki:2006we}, the extension of theoretical methods to enable calculations at large $\mu_B$ remains a challenge. Extensive efforts are underway to search for a possible critical point in the QCD phase diagram. 

Fluctuations of event-by-event net-proton number in nuclear collision experiments have been suggested as sensitive probes to search for the QCD critical point (CP)~\cite{Stephanov:2008qz,Asakawa:2009aj,Stephanov:2011pb}. The fluctuations of a variable are often quantified via cumulants. Cumulants of net-proton number ($N$) up to fourth order can be expressed as: $C_1 = \langle N \rangle$, $C_2 = \langle \delta N^2 \rangle$,  $C_3 = \langle \delta N^3 \rangle$, and  $C_4 = \langle \delta N^4 \rangle -3 \langle \delta N^2 \rangle$, where $\delta N = N - \langle N \rangle$. These cumulants are related to the correlation length ($\xi$) of the system formed in heavy-ion collisions~\cite{Stephanov:2011pb}, which diverges at a critical point. Ratios of cumulants can be directly compared with ratios of thermodynamic number susceptibilities obtained in various theory calculations~\cite{Gavai:2010zn,Gupta:2011wh,Karsch:2010ck,Garg:2013ata}. A non-monotonic dependence on collision energy in the ratio of the fourth- to the second-order net-proton cumulant ($C_4/C_2$, also often referred to as the moment product $\kappa\sigma^2$, where $\kappa$ is the kurtosis and $\sigma^2$ is the variance~\cite{STAR:2021iop}) relative to baseline fluctuations has been proposed by a model-based study as the signature of the CP~\cite{Stephanov:2011pb}. In the first phase of the Beam Energy Scan (BES-I) program at RHIC, the Solenoidal Tracker at RHIC (STAR) experiment extensively analyzed net-proton fluctuations~\cite{STAR:2010mib,STAR:2013gus,STAR:2020tga,STAR:2021iop,STAR:2021rls,STAR:2022vlo}.
However, precision measurements were needed to reduce the large uncertainties on the first set of results~\cite{STAR:2020tga} to draw concrete conclusions. Recent $C_4/C_2$ results from STAR’s fixed target (FXT) Au+Au collisions at $\sqrt{s_{NN}}=3$ GeV show suppressed values relative to higher collision energies and are consistent with a hadronic transport model without a CP~\cite{STAR:2021fge,STAR:2022etb}. A second phase of the Beam Energy Scan (BES-II) with large event statistics at $\sqrt{s_{NN}}$ = 7.7 -- 27 GeV and important detector upgrades, including the replacement of the inner sectors of the Time Projection Chamber (iTPC)~\cite{ref_iTPC_BESII}, has recently been concluded.

This letter presents the latest results on net-proton cumulants up to fourth order from Au+Au collisions at $\sqrt{s_{NN}}=7.7, 9.2, 11.5, 14.6, 17.3, 19.6$ and 27 GeV from the BES-II program. All datasets were taken in collider mode. The two energies $\sqrt{s_{NN}}=$ 9.2 and 17.3 GeV were not present in BES-I. 
Furthermore, proton factorial cumulants ($\kappa_n$), also suggested as promising probes for CP search~\cite{Bzdak:2016sxg}, are reported up to the fourth order. They are defined as $\kappa_1 = C_1$, $\kappa_2 = -C_1 + C_2$, $\kappa_3 = 2C_1 - 3C_2 + C_3$ and $\kappa_4 =-6C_1+11C_2-6C_3+C_4$. The factorial cumulant of a certain order reflects genuine multiparticle correlation of that order whereas the regular cumulant of that order also contains contributions from lower orders.

Au+Au collisions resulting in signals in the trigger detectors above the noise threshold level (minimum bias) are analyzed~\cite{Adler:2000bd,Llope:2003ti}. The $z$-coordinate of the interaction point (also called the primary vertex), is denoted $V_z$, and is constrained to be within $\pm50$ cm from the center of the STAR detector at all energies except 27 GeV where it is $\pm27$ cm to ensure uniform acceptance and efficiency of the TPC. This choice of $V_z$ cut at  $\sqrt{s_{NN}}=27$ GeV is due to reduced TPC acceptance as the iTPC upgrade had not taken place by the time of data taking. The total numbers of analyzed events are listed in Table~\ref{tab1_stats}, and are about $7-18$ times larger than available in BES-I~\cite{STAR:2021iop}. 
(Anti-)protons are identified within a rapidity acceptance of $|y|<0.5$ and transverse momentum range of $p_T$ = 0.4 to 2.0 GeV/$c$  based on their ionization energy loss using the TPC. Along with the TPC, the Time-of-Flight detector (TOF) is used from $p_T$ = 0.8 to 2.0 GeV/$c$~\cite{STAR:2002eio}. This ensures a high purity, $\sim99\%$ over the full $y$ -- $p_T$ phase space. The distance of the closest approach of a track to the primary vertex, called the DCA, is taken to be less than 1 cm to suppress contributions from secondaries~\cite{STAR:2013gus} and weak decays~\cite{STAR:2001rbj}. Moreover, the weak decays only mildly affect the cumulant ratios~\cite{Vovchenko:2021kxx,Zhang:2019lqz}. 

\begin{table}
	\caption{Total event statistics (in millions) in Au+Au collisions for various collision energies ($\sqrt{s_{NN}}$) in BES-II.}
	\centering   
	\begin{tabular}{|c|c|c|c|c|c|c|c|}
		\hline
		~$\sqrt{s_{NN}}$ (GeV)~ & ~7.7~ & ~9.2~ & ~11.5~ & ~14.6~ & ~17.3~ & ~19.6~ & ~27~  \\
		\hline
		Events ($10^6$)& 45 & 78 & 116 & 178 & 116 & 270 & 220 \\
		\hline
	\end{tabular}
	\label{tab1_stats}
\end{table}

\begin{figure}[!htbp]
	\centering
	\includegraphics[width=\linewidth]{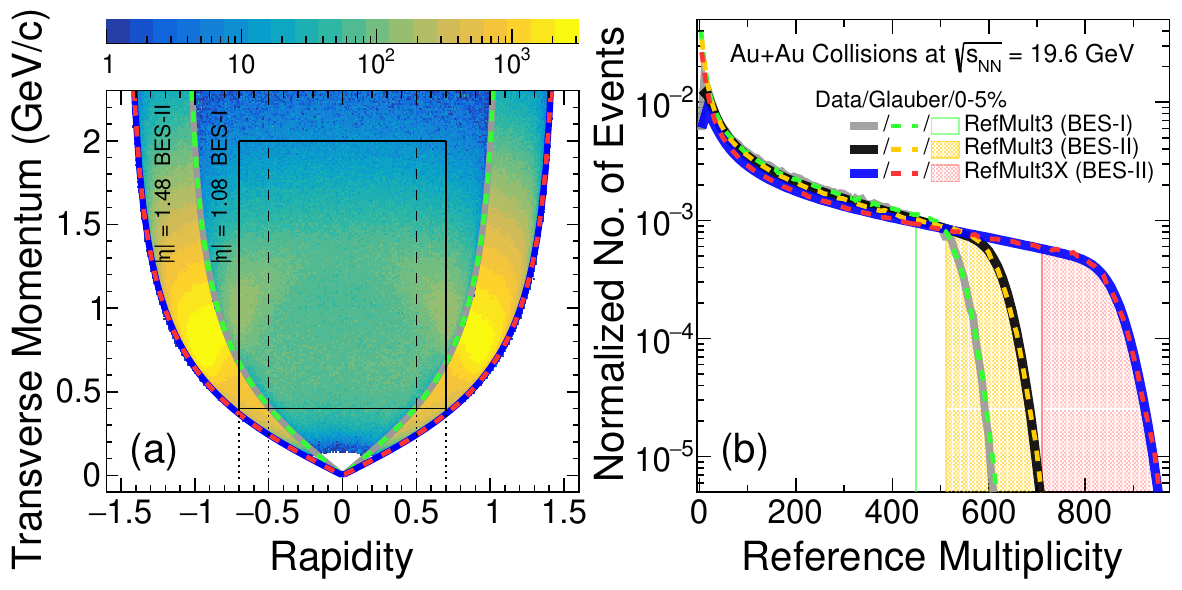}
	\caption{Panel (a): (Anti)proton acceptances from the TPC in transverse momentum ($p_T$) vs. rapidity ($y$) phase space in Au+Au collisions at $\sqrt{s_{\text{NN}}}$ = 19.6 GeV from BES-II and BES-I. The green and red dashed curves describe the pseudorapidity ($\eta$) boundaries of the acceptances in BES-I ($|\eta|=1.08$) and BES-II ($|\eta|=1.48$), respectively. Only BES-I acceptance is plotted in the common region. (Anti)protons are selected in the 0--5\% most central events within $13<V_{z}<15$ cm for positive rapidities and $-15<V_{z}<-13$ cm for negative rapidities. These $V_z$ ranges are taken just as an example. The dashed box denotes the acceptance criterion ($|y|<0.5$) for this analysis, which is the same as BES-I. Panel (b): Reference charged particle multiplicity distributions RefMult3 (BES-I green line and BES-II black line) and RefMult3X (BES-II blue line). Distributions from data and Glauber fits are presented. The shaded bands reflect the 0--5\% central collision events.}
	\label{labl_fig1}
\end{figure}

Fig.~\ref{labl_fig1} (a) presents the $y$ -- $p_T$ acceptance of (anti)proton yields identified by the TPC in Au+Au collisions at $\sqrt{s_{\text{NN}}}$ = 19.6 GeV from BES-I (smaller band) and BES-II (larger band). The acceptance for the current analysis ($|y|<0.5$ and $0.4<p_T<2.0$ GeV/c) is the same as that used in BES-I~\cite{STAR:2020tga} (shown as a dashed box in Fig.~\ref{labl_fig1} (a)). The iTPC upgrade in BES-II highly extends the acceptance, which provides a possibility of measurement over $|y|<0.7$ in the future. Fig.~\ref{labl_fig1} (b) shows the reference charged particle multiplicity distributions from data: RefMult3 and RefMult3X used for defining collision centrality. They are defined by the  charged particle multiplicities, not corrected for efficiency,  within $|\eta|<1.0$ and $|\eta|<1.6$, respectively, after excluding protons and antiprotons to avoid the self-correlation effect~\cite{Chatterjee:2019fey,Luo:2017faz}. Due to the increased efficiency from the iTPC, RefMult3 is larger in BES-II than it was in BES-I.  RefMult3X is larger still, due to the larger $\eta$ acceptance.  Larger multiplicity ranges lead to better centrality resolution. The RefMult3 and RefMult3X distributions are fitted with multiplicity obtained from a Glauber model simulation to extract the number of participating nucleons ($N_{\rm part}$) and determine the centrality cuts. An event-weighted averaging called the centrality-bin-width correction (CBWC) was performed on the measurements for a given centrality~\cite{Luo:2013bmi} (see Fig. 1 of the supplemental material for details~\cite{ref_suppl_material}). The measurements are corrected for finite detector efficiency by assuming the detector response to be binomial~\cite{Bzdak:2012ab,Kitazawa:2012at,Luo:2014rea,Nonaka:2017kko,Luo:2018ofd}. The track selection cuts, the background estimates, and detector efficiency are varied for evaluating systematic uncertainties on the measurements. All sources of systematics are subjected to a Barlow check~\cite{Barlow:2002yb} and are added in quadrature to obtain total systematic uncertainties ($\sigma_{\rm sys}$). For net-proton $C_4/C_2$ measurements, the PID and DCA variations are dominant contributors. As an example, at $\sqrt{s_{NN}}$ = 7.7 GeV, PID and DCA contribution are 88\% and 47\% of $\sigma_{\rm sys}$ for 0-5\% centrality, respectively. Enhanced event statistics and better tracking and PID from iTPC in BES-II lead to a significant reduction of uncertainties. A factor of 4.7 (3.2) reduction in statistical (systematic) uncertainties on net-proton $C_4/C_2$ in the central collisions from BES-II has been achieved compared to that in BES-I. 
\begin{figure*}[!htbp]
	\centering
	\includegraphics[width=\linewidth]{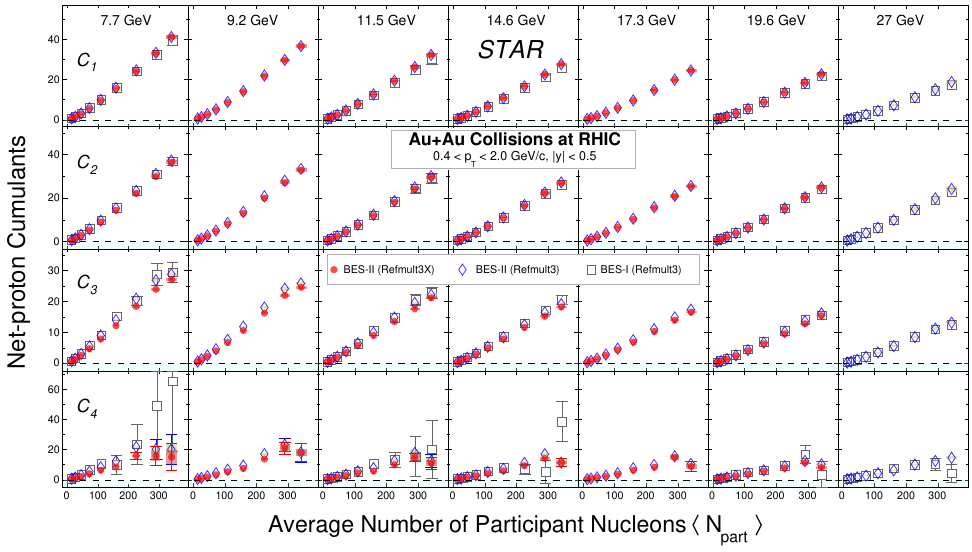}
	\caption{Cumulants of net-proton multiplicity distribution from $\sqrt{s_{NN}}$ = 7.7 -- 27 GeV as a function of collision centrality ($\langle N_{\rm part} \rangle$; average number of participating nucleons) in Au+Au collisions. (Anti-) protons are measured at mid-rapidity ($|y|<0.5$) within $0.4 < p_T < 2.0$ GeV/$c$. Results from BES-II with RefMult3X (RefMult3) used for centrality definition are shown as red (blue) markers, while those from BES-I~\cite{STAR:2020tga,STAR:2021iop} (RefMult3) are shown as open squares. The bars and bands on the data points from BES-II represent statistical and systematic uncertainties, respectively. Total uncertainties are shown for BES-I data points as bars on data points.}
	\label{labl_fig2}
\end{figure*}

Net-proton cumulants ($C_n$) up to fourth order from BES-II as a function of collision centrality are shown in Fig.~\ref{labl_fig2} along with those from BES-I. In general, the cumulants smoothly increase as a function of centrality and decrease from lower to higher collision energies. 
Higher-order cumulants measured using RefMult3X for centrality are lower than those using RefMult3, and those using RefMult3 in BES-II are lower than those seen in BES-I, though mostly consistent within uncertainties. Such a trend is expected due to the differences in centrality resolution~\cite{Luo:2013bmi} (see the supplemental material for more discussion~\cite{ref_suppl_material}). Factorial cumulants, being a combination of regular cumulants, also show such dependence on resolution.
Note that due to the absence of the iTPC during recording of $\sqrt{s_{NN}}$ = 27 GeV Au+Au collisions from BES-II, only results using RefMult3 are presented at this energy. $C_4$ measurements as well as the cumulant ratio $C_4/C_2$ in 0-5\% collisions are consistent across different centrality measures, and are consistent with BES-I data (see Fig. 4 of supplemental material for details~\cite{ref_suppl_material}).

\begin{figure*}[!htbp]
	\centering	
	\includegraphics[width=\linewidth]{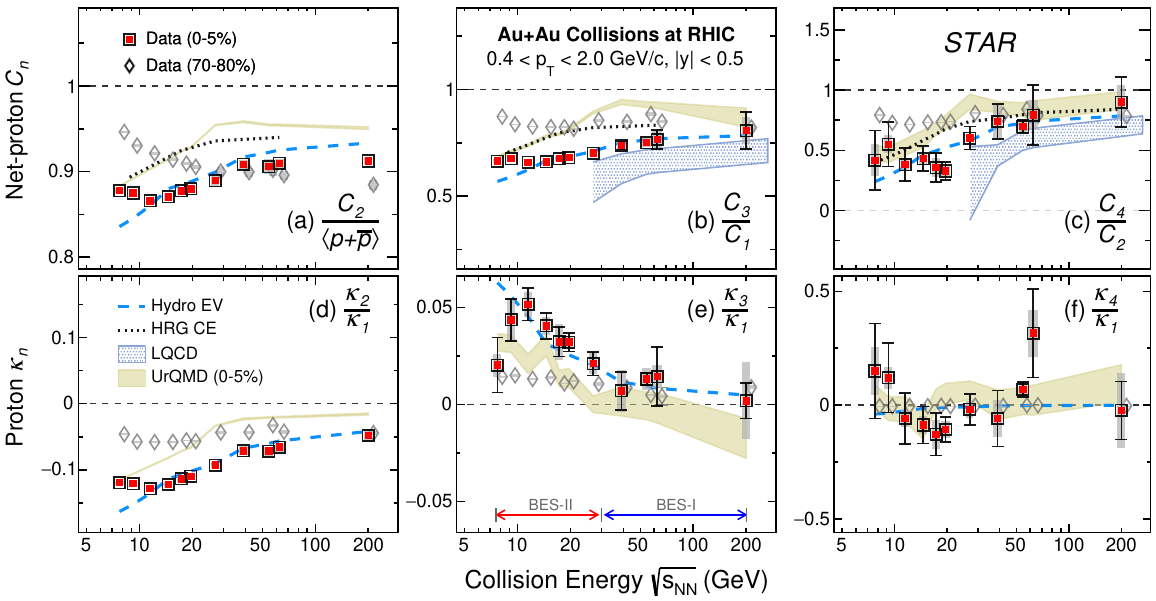}
	\caption{Net-proton cumulant ratios: (a) $C_2/\langle p+\bar{p}\rangle$, (b) $C_3/C_1$, and (c) $C_4/C_2$ and proton factorial cumulant ratios: (d) $\kappa_2/\kappa_1$, (e) $\kappa_3/\kappa_1$ and (f) $\kappa_4/\kappa_1$ in Au+Au collisions. Results from  BES-II ($\sqrt{s_{NN}}$ = 7.7 -- 27 GeV with RefMult3X) and BES-I~\cite{STAR:2020tga,STAR:2021iop} ($\sqrt{s_{NN}}$ = 39 -- 200 GeV with RefMult3) program at RHIC are shown. (Anti-)protons are measured at mid-rapidity ($|y|<0.5$) within $0.4 < p_T < 2.0$ GeV/$c$. The bars and bands on the data points reflect statistical and systematic uncertainties, respectively. Theoretical calculations from a hydrodynamical model~\cite{Vovchenko:2021kxx} (Hydro, blue dashed line), thermal model with canonical treatment for baryon charge~\cite{Braun-Munzinger:2020jbk} (HRG CE, black dashed line), transport model~\cite{Bass:1998ca,Bleicher:1999xi} (UrQMD, brown band), and lattice QCD~\cite{Bazavov:2020bjn,Bollweg:2024epj} (LQCD, light blue band) are also presented.}
	\label{labl_fig3}
\end{figure*}

Figure~\ref{labl_fig3} shows the collision energy dependence of ratios of net-proton cumulants and proton factorial cumulants from BES-II and BES-I for the most central 0-5\% and peripheral 70-80\% collisions.  BES-II results shown are measured with RefMult3X for centrality definition except for $\sqrt{s_{NN}}$ = 27 GeV. By construction, the Poisson baseline for net-proton cumulant ratios is at unity while for proton factorial cumulants it is at zero. A strong deviation from the Poisson baseline is observed in all ratios. The cumulant ratios for 0-5\% centrality decrease with decreasing collision energy except at low energies where hint of a possible rise can be seen, albiet with significant uncertainties. As mentioned earlier, the factorial cumulants reflect genuine multiparticle correlation. In the case of proton $\kappa_2/\kappa_1$ and $\kappa_4/\kappa_1$, for 0-5\% central collisions, a similar trend in collision energy dependence is observed as seen for the net-proton cumulant ratios in the same collision centrality. For $\kappa_3/\kappa_1$ (0-5\%), from high to low collision energy, an initial increase followed by a dip towards the Poisson baseline is observed. The peripheral 70-80\% collision data are closer to the Poisson baseline at zero compared to 0-5\% data and for $\kappa_4/\kappa_1$ it is consistent with zero at all energies. 

\begin{figure}[!htbp]
	\centering
	\includegraphics[width=\linewidth]{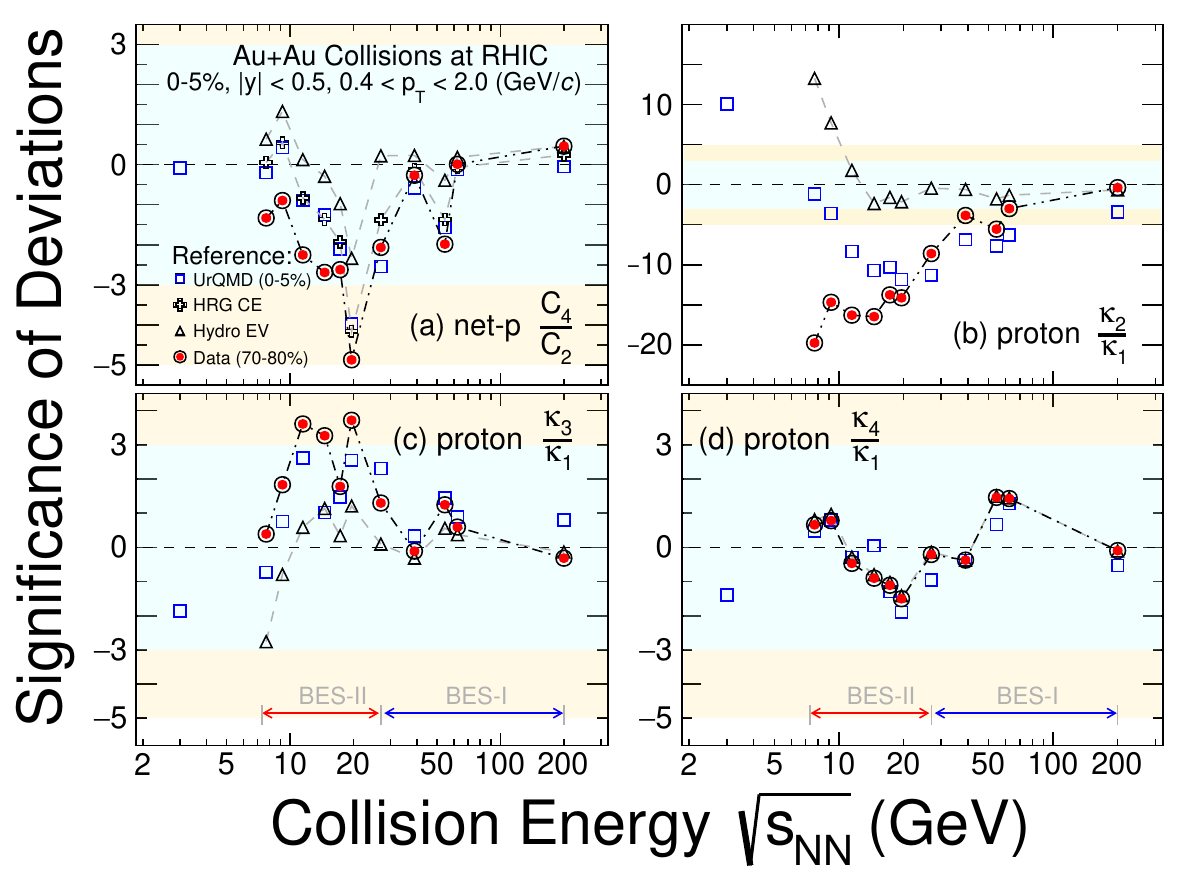}
	\caption{Significance of deviation (data$-$reference)/$\sigma_{\rm total}$ for (a) net-proton cumulant ratios $C_4/C_2$; proton factorial cumulant ratios (b) $\kappa_4/\kappa_1$; (c) $\kappa_3/\kappa_1$ and (d) $\kappa_2/\kappa_1$ in 0-5\% Au+Au collisions~\cite{STAR:2020tga,STAR:2021iop,STAR:2021fge,STAR:2022etb}. References include the non-critical model calculations, such as the UrQMD transport model~\cite{Bass:1998ca,Bleicher:1999xi} (blue square), HRG with canonical ensemble for baryon charge~\cite{Braun-Munzinger:2020jbk} (HRG CE, black cross), hydrodynamic model with excluded volume~\cite{Vovchenko:2021kxx} (Hydro EV, black triangle), and data from 70-80\% peripheral collisions (red dots).}
	\label{labl_fig4}
\end{figure}

The data are compared to calculations without a critical point, such as hadronic transport model UrQMD~\cite{Bass:1998ca,Bleicher:1999xi}, a thermal model with canonical treatment of baryon charge (HRG CE)~\cite{Braun-Munzinger:2020jbk}, and a hydrodynamic model with excluded volume (hydro EV)~\cite{Vovchenko:2021kxx}. The UrQMD results are analyzed in the same way as the experimental data. None of the models fully describes the collision energy dependence of the data. These are distinct theoretical models describing the dynamics of QGP evolution and all of them incorporate exact baryon number conservation, an important aspect to be considered, especially at low collision energies.
Based on BES-I, possible evidence of a non-monotonic trend in 0-5\% $C_4/C_2$ with respect to the Poisson baseline was reported~\cite{STAR:2020tga}. However, the precision data from BES-II do not support such a trend, and always remain below the Poisson baseline. The new data, although closer to the non-critical model calculations with baryon number conservation than to Poisson expectations, exhibit a deviation from the former at around 20 GeV (see more discussion in Fig.~\ref{labl_fig4}). 

STAR has previously reported proton $C_4/C_2$ measurement at $\sqrt{s_{NN}}=3$ GeV with the rapidity acceptance of $-0.5<y<0$ from its FXT program~\cite{STAR:2021fge,STAR:2022etb}. Good agreement between data and UrQMD had been observed. In addition, data for $\sqrt{s_{NN}}>27$ GeV are found to be consistent with these non-CP model calculations and lattice-QCD (LQCD) calculations with grand canonical ensemble~\cite{Bazavov:2020bjn}. LQCD studies indicate no sign of a critical point corresponding to chemical freezeout points ($\mu_B$,T) for $\sqrt{s_{NN}}>27$ GeV and instead suggest a crossover in this regime.

The deviation between experimental data and non-critical baselines/references is shown in Fig.~\ref{labl_fig4} for net-proton $C_4/C_2$ (a), proton $\kappa_2/\kappa_1$ (b), proton $\kappa_3/\kappa_1$ (c), and proton $\kappa_4/\kappa_1$ (d). These deviations are obtained by taking the difference between the 0-5\% data and baselines and dividing with the total uncertainties ($\sigma_{\rm total}$, obtained adding uncertainties in data and baselines in quadrature). Three typical calculations including the UrQMD, the HRG CE, and the hydro EV are used in the analysis. In addition, the 70-80\% peripheral collision data are used for comparison.

For the net-proton cumulant ratio $C_4/C_2$, maximum deviations of 2 -- 5$\sigma$ are seen at $\sqrt{s_{NN}}$ = 19.6 GeV from all references. A minimum at same collision energy is also seen for data after subtracting the references (see Fig. 5 of the supplemental material~\cite{ref_suppl_material}). On the other hand, in the case of collisions at 3 GeV~\cite{STAR:2021fge} or above 27 GeV~\cite{STAR:2020tga}, the central data is consistent with all of the references within $\sim2\sigma$. For the factorial cumulants ratios, the amplitude of the deviation seems to decrease as a function of the order of the correlation: the maximum deviation is seen in $\kappa_2/\kappa_1$ while the minimum is in $\kappa_4/\kappa_1$, shown in Fig.~\ref{labl_fig4} (b) and (d), respectively. Clearly, precise experimental data between $\sqrt{s_{NN}}$ = 3.0 and 7.7 GeV is needed to extend the search for the signal of the QCD critical point and the 1st-order phase boundary at the high baryon density region. The exploration of this region with precision measurements is currently underway using STAR's FXT program.

In summary, new high precision data on net-proton cumulants and proton factorial cumulants and their ratios from the second phase of RHIC beam energy scan ($\sqrt{s_{NN}}$ = 7.7 -- 27 GeV, corresponding to $\mu_B$ = 400 -- 150 MeV~\cite{STAR:2017sal}), are reported in this paper. Along with improved statistics, the RHIC BES-II program provides better centrality resolution and better control on the systematic uncertainties of the measurements. New net-proton $C_4/C_2$ data in 0-5\% Au+Au collisions are consistent with those from BES-I. They remain below the Poisson baseline at unity at all collision energies. The new results compared to non-critical model calculations and peripheral 70-80\% data show a minimum for significance of deviation around 19.6 GeV at $\sim2$ -- $5\sigma$. The highest (lowest) significance level of $5\sigma$ ($2\sigma$) deviation corresponds to the case of peripheral collision data (hydro EV calculations) taken as the baseline. Deviations are also seen for proton factorial cumulant ratios, with higher significance for $\kappa_2/\kappa_1$ and $\kappa_3/\kappa_1$ compared to $\kappa_4/\kappa_1$. As it has been argued~\cite{Stephanov:2011pb}, the minimum seen in net-proton $C_4/C_2$ data is a characteristic feature of the proposed signature of the critical point for strongly interacting matter. 
Dynamical model calculations including the criticality are needed to fully understand the data. These data will be crucial for constraining the equation of state~\cite{An:2021wof} of the medium created in these collisions, . 

We thank the RHIC Operations Group and RCF at BNL, the NERSC Center at LBNL, and the Open Science Grid consortium for providing resources and support. This work was supported in part by the Office of Nuclear Physics within the U.S. DOE Office of Science, the U.S. National Science Foundation, National Natural Science Foundation of China, Chinese Academy of Science, the Ministry of Science and Technology of China and the Chinese Ministry of Education, the Higher Education Sprout Project by Ministry of Education at NCKU, the National Research Foundation of Korea, Czech Science Foundation and Ministry of Education, Youth and Sports of the Czech Republic, Hungarian National Research, Development and Innovation Office, New National Excellency Programme of the Hungarian Ministry of Human Capacities, Department of Atomic Energy and Department of Science and Technology of the Government of India, the National Science Centre and WUT ID-UB of Poland, the Ministry of Science, Education and Sports of the Republic of Croatia, German Bundesministerium f\"ur Bildung, Wissenschaft, Forschung and Technologie (BMBF), Helmholtz Association, Ministry of Education, Culture, Sports, Science, and Technology (MEXT), Japan Society for the Promotion of Science (JSPS) and Agencia Nacional de Investigaci\'on y Desarrollo (ANID) of Chile.






\bibliographystyle{apsrev4-1} 
\bibliography{References_list} 

\begin{thebibliography}{44}%
\makeatletter
\providecommand \@ifxundefined [1]{%
 \@ifx{#1\undefined}
}%
\providecommand \@ifnum [1]{%
 \ifnum #1\expandafter \@firstoftwo
 \else \expandafter \@secondoftwo
 \fi
}%
\providecommand \@ifx [1]{%
 \ifx #1\expandafter \@firstoftwo
 \else \expandafter \@secondoftwo
 \fi
}%
\providecommand \natexlab [1]{#1}%
\providecommand \enquote  [1]{``#1''}%
\providecommand \bibnamefont  [1]{#1}%
\providecommand \bibfnamefont [1]{#1}%
\providecommand \citenamefont [1]{#1}%
\providecommand \href@noop [0]{\@secondoftwo}%
\providecommand \href [0]{\begingroup \@sanitize@url \@href}%
\providecommand \@href[1]{\@@startlink{#1}\@@href}%
\providecommand \@@href[1]{\endgroup#1\@@endlink}%
\providecommand \@sanitize@url [0]{\catcode `\\12\catcode `\$12\catcode
  `\&12\catcode `\#12\catcode `\^12\catcode `\_12\catcode `\%12\relax}%
\providecommand \@@startlink[1]{}%
\providecommand \@@endlink[0]{}%
\providecommand \url  [0]{\begingroup\@sanitize@url \@url }%
\providecommand \@url [1]{\endgroup\@href {#1}{\urlprefix }}%
\providecommand \urlprefix  [0]{URL }%
\providecommand \Eprint [0]{\href }%
\providecommand \doibase [0]{http://dx.doi.org/}%
\providecommand \selectlanguage [0]{\@gobble}%
\providecommand \bibinfo  [0]{\@secondoftwo}%
\providecommand \bibfield  [0]{\@secondoftwo}%
\providecommand \translation [1]{[#1]}%
\providecommand \BibitemOpen [0]{}%
\providecommand \bibitemStop [0]{}%
\providecommand \bibitemNoStop [0]{.\EOS\space}%
\providecommand \EOS [0]{\spacefactor3000\relax}%
\providecommand \BibitemShut  [1]{\csname bibitem#1\endcsname}%
\let\auto@bib@innerbib\@empty
\bibitem [{\citenamefont {Rajagopal}\ and\ \citenamefont
  {Wilczek}(2000)}]{Rajagopal:2000wf}%
  \BibitemOpen
  \bibfield  {author} {\bibinfo {author} {\bibfnamefont {K.}~\bibnamefont
  {Rajagopal}}\ and\ \bibinfo {author} {\bibfnamefont {F.}~\bibnamefont
  {Wilczek}},\ }\enquote {\bibinfo {title} {{The Condensed matter physics of
  QCD}},}\ in\ \href {\doibase 10.1142/9789812810458_0043} {\emph {\bibinfo
  {booktitle} {{At the frontier of particle physics. Handbook of QCD. Vol.
  1-3}}}},\ \bibinfo {editor} {edited by\ \bibinfo {editor} {\bibfnamefont
  {M.}~\bibnamefont {Shifman}}\ and\ \bibinfo {editor} {\bibfnamefont
  {B.}~\bibnamefont {Ioffe}}}\ (\bibinfo {year} {2000})\ pp.\ \bibinfo {pages}
  {2061--2151},\ \Eprint {http://arxiv.org/abs/hep-ph/0011333}
  {arXiv:hep-ph/0011333} \BibitemShut {NoStop}%
\bibitem [{\citenamefont {Bzdak}\ \emph {et~al.}(2020)\citenamefont {Bzdak},
  \citenamefont {Esumi}, \citenamefont {Koch}, \citenamefont {Liao},
  \citenamefont {Stephanov},\ and\ \citenamefont {Xu}}]{Bzdak:2019pkr}%
  \BibitemOpen
  \bibfield  {author} {\bibinfo {author} {\bibfnamefont {A.}~\bibnamefont
  {Bzdak}}, \bibinfo {author} {\bibfnamefont {S.}~\bibnamefont {Esumi}},
  \bibinfo {author} {\bibfnamefont {V.}~\bibnamefont {Koch}}, \bibinfo {author}
  {\bibfnamefont {J.}~\bibnamefont {Liao}}, \bibinfo {author} {\bibfnamefont
  {M.}~\bibnamefont {Stephanov}}, \ and\ \bibinfo {author} {\bibfnamefont
  {N.}~\bibnamefont {Xu}},\ }\href {\doibase 10.1016/j.physrep.2020.01.005}
  {\bibfield  {journal} {\bibinfo  {journal} {Phys. Rept.}\ }\textbf {\bibinfo
  {volume} {853}},\ \bibinfo {pages} {1} (\bibinfo {year} {2020})},\ \Eprint
  {http://arxiv.org/abs/1906.00936} {arXiv:1906.00936 [nucl-th]} \BibitemShut
  {NoStop}%
\bibitem [{\citenamefont {Pandav}\ \emph {et~al.}(2022)\citenamefont {Pandav},
  \citenamefont {Mallick},\ and\ \citenamefont {Mohanty}}]{Pandav:2022xxx}%
  \BibitemOpen
  \bibfield  {author} {\bibinfo {author} {\bibfnamefont {A.}~\bibnamefont
  {Pandav}}, \bibinfo {author} {\bibfnamefont {D.}~\bibnamefont {Mallick}}, \
  and\ \bibinfo {author} {\bibfnamefont {B.}~\bibnamefont {Mohanty}},\ }\href
  {\doibase 10.1016/j.ppnp.2022.103960} {\bibfield  {journal} {\bibinfo
  {journal} {Prog. Part. Nucl. Phys.}\ }\textbf {\bibinfo {volume} {125}},\
  \bibinfo {pages} {103960} (\bibinfo {year} {2022})},\ \Eprint
  {http://arxiv.org/abs/2203.07817} {arXiv:2203.07817 [nucl-ex]} \BibitemShut
  {NoStop}%
\bibitem [{\citenamefont {Aoki}\ \emph {et~al.}(2006)\citenamefont {Aoki},
  \citenamefont {Endrodi}, \citenamefont {Fodor}, \citenamefont {Katz},\ and\
  \citenamefont {Szabo}}]{Aoki:2006we}%
  \BibitemOpen
  \bibfield  {author} {\bibinfo {author} {\bibfnamefont {Y.}~\bibnamefont
  {Aoki}}, \bibinfo {author} {\bibfnamefont {G.}~\bibnamefont {Endrodi}},
  \bibinfo {author} {\bibfnamefont {Z.}~\bibnamefont {Fodor}}, \bibinfo
  {author} {\bibfnamefont {S.~D.}\ \bibnamefont {Katz}}, \ and\ \bibinfo
  {author} {\bibfnamefont {K.~K.}\ \bibnamefont {Szabo}},\ }\href {\doibase
  10.1038/nature05120} {\bibfield  {journal} {\bibinfo  {journal} {Nature}\
  }\textbf {\bibinfo {volume} {443}},\ \bibinfo {pages} {675} (\bibinfo {year}
  {2006})},\ \Eprint {http://arxiv.org/abs/hep-lat/0611014}
  {arXiv:hep-lat/0611014} \BibitemShut {NoStop}%
\bibitem [{\citenamefont {Stephanov}(2009)}]{Stephanov:2008qz}%
  \BibitemOpen
  \bibfield  {author} {\bibinfo {author} {\bibfnamefont {M.~A.}\ \bibnamefont
  {Stephanov}},\ }\href {\doibase 10.1103/PhysRevLett.102.032301} {\bibfield
  {journal} {\bibinfo  {journal} {Phys. Rev. Lett.}\ }\textbf {\bibinfo
  {volume} {102}},\ \bibinfo {pages} {032301} (\bibinfo {year} {2009})},\
  \Eprint {http://arxiv.org/abs/0809.3450} {arXiv:0809.3450 [hep-ph]}
  \BibitemShut {NoStop}%
\bibitem [{\citenamefont {Asakawa}\ \emph {et~al.}(2009)\citenamefont
  {Asakawa}, \citenamefont {Ejiri},\ and\ \citenamefont
  {Kitazawa}}]{Asakawa:2009aj}%
  \BibitemOpen
  \bibfield  {author} {\bibinfo {author} {\bibfnamefont {M.}~\bibnamefont
  {Asakawa}}, \bibinfo {author} {\bibfnamefont {S.}~\bibnamefont {Ejiri}}, \
  and\ \bibinfo {author} {\bibfnamefont {M.}~\bibnamefont {Kitazawa}},\ }\href
  {\doibase 10.1103/PhysRevLett.103.262301} {\bibfield  {journal} {\bibinfo
  {journal} {Phys. Rev. Lett.}\ }\textbf {\bibinfo {volume} {103}},\ \bibinfo
  {pages} {262301} (\bibinfo {year} {2009})},\ \Eprint
  {http://arxiv.org/abs/0904.2089} {arXiv:0904.2089 [nucl-th]} \BibitemShut
  {NoStop}%
\bibitem [{\citenamefont {Stephanov}(2011)}]{Stephanov:2011pb}%
  \BibitemOpen
  \bibfield  {author} {\bibinfo {author} {\bibfnamefont {M.~A.}\ \bibnamefont
  {Stephanov}},\ }\href {\doibase 10.1103/PhysRevLett.107.052301} {\bibfield
  {journal} {\bibinfo  {journal} {Phys. Rev. Lett.}\ }\textbf {\bibinfo
  {volume} {107}},\ \bibinfo {pages} {052301} (\bibinfo {year} {2011})},\
  \Eprint {http://arxiv.org/abs/1104.1627} {arXiv:1104.1627 [hep-ph]}
  \BibitemShut {NoStop}%
\bibitem [{\citenamefont {Gavai}\ and\ \citenamefont
  {Gupta}(2011)}]{Gavai:2010zn}%
  \BibitemOpen
  \bibfield  {author} {\bibinfo {author} {\bibfnamefont {R.~V.}\ \bibnamefont
  {Gavai}}\ and\ \bibinfo {author} {\bibfnamefont {S.}~\bibnamefont {Gupta}},\
  }\href {\doibase 10.1016/j.physletb.2011.01.006} {\bibfield  {journal}
  {\bibinfo  {journal} {Phys. Lett. B}\ }\textbf {\bibinfo {volume} {696}},\
  \bibinfo {pages} {459} (\bibinfo {year} {2011})},\ \Eprint
  {http://arxiv.org/abs/1001.3796} {arXiv:1001.3796 [hep-lat]} \BibitemShut
  {NoStop}%
\bibitem [{\citenamefont {Gupta}\ \emph {et~al.}(2011)\citenamefont {Gupta},
  \citenamefont {Luo}, \citenamefont {Mohanty}, \citenamefont {Ritter},\ and\
  \citenamefont {Xu}}]{Gupta:2011wh}%
  \BibitemOpen
  \bibfield  {author} {\bibinfo {author} {\bibfnamefont {S.}~\bibnamefont
  {Gupta}}, \bibinfo {author} {\bibfnamefont {X.}~\bibnamefont {Luo}}, \bibinfo
  {author} {\bibfnamefont {B.}~\bibnamefont {Mohanty}}, \bibinfo {author}
  {\bibfnamefont {H.~G.}\ \bibnamefont {Ritter}}, \ and\ \bibinfo {author}
  {\bibfnamefont {N.}~\bibnamefont {Xu}},\ }\href {\doibase
  10.1126/science.1204621} {\bibfield  {journal} {\bibinfo  {journal}
  {Science}\ }\textbf {\bibinfo {volume} {332}},\ \bibinfo {pages} {1525}
  (\bibinfo {year} {2011})},\ \Eprint {http://arxiv.org/abs/1105.3934}
  {arXiv:1105.3934 [hep-ph]} \BibitemShut {NoStop}%
\bibitem [{\citenamefont {Karsch}\ and\ \citenamefont
  {Redlich}(2011)}]{Karsch:2010ck}%
  \BibitemOpen
  \bibfield  {author} {\bibinfo {author} {\bibfnamefont {F.}~\bibnamefont
  {Karsch}}\ and\ \bibinfo {author} {\bibfnamefont {K.}~\bibnamefont
  {Redlich}},\ }\href {\doibase 10.1016/j.physletb.2010.10.046} {\bibfield
  {journal} {\bibinfo  {journal} {Phys. Lett. B}\ }\textbf {\bibinfo {volume}
  {695}},\ \bibinfo {pages} {136} (\bibinfo {year} {2011})},\ \Eprint
  {http://arxiv.org/abs/1007.2581} {arXiv:1007.2581 [hep-ph]} \BibitemShut
  {NoStop}%
\bibitem [{\citenamefont {Garg}\ \emph {et~al.}(2013)\citenamefont {Garg},
  \citenamefont {Mishra}, \citenamefont {Netrakanti}, \citenamefont {Mohanty},
  \citenamefont {Mohanty}, \citenamefont {Singh},\ and\ \citenamefont
  {Xu}}]{Garg:2013ata}%
  \BibitemOpen
  \bibfield  {author} {\bibinfo {author} {\bibfnamefont {P.}~\bibnamefont
  {Garg}}, \bibinfo {author} {\bibfnamefont {D.~K.}\ \bibnamefont {Mishra}},
  \bibinfo {author} {\bibfnamefont {P.~K.}\ \bibnamefont {Netrakanti}},
  \bibinfo {author} {\bibfnamefont {B.}~\bibnamefont {Mohanty}}, \bibinfo
  {author} {\bibfnamefont {A.~K.}\ \bibnamefont {Mohanty}}, \bibinfo {author}
  {\bibfnamefont {B.~K.}\ \bibnamefont {Singh}}, \ and\ \bibinfo {author}
  {\bibfnamefont {N.}~\bibnamefont {Xu}},\ }\href {\doibase
  10.1016/j.physletb.2013.09.019} {\bibfield  {journal} {\bibinfo  {journal}
  {Phys. Lett. B}\ }\textbf {\bibinfo {volume} {726}},\ \bibinfo {pages} {691}
  (\bibinfo {year} {2013})},\ \Eprint {http://arxiv.org/abs/1304.7133}
  {arXiv:1304.7133 [nucl-ex]} \BibitemShut {NoStop}%
\bibitem [{\citenamefont {Abdallah}\ \emph
  {et~al.}(2021{\natexlab{a}})\citenamefont {Abdallah} \emph
  {et~al.}}]{STAR:2021iop}%
  \BibitemOpen
  \bibfield  {author} {\bibinfo {author} {\bibfnamefont {M.}~\bibnamefont
  {Abdallah}} \emph {et~al.} (\bibinfo {collaboration} {STAR}),\ }\href
  {\doibase 10.1103/PhysRevC.104.024902} {\bibfield  {journal} {\bibinfo
  {journal} {Phys. Rev. C}\ }\textbf {\bibinfo {volume} {104}},\ \bibinfo
  {pages} {024902} (\bibinfo {year} {2021}{\natexlab{a}})},\ \bibinfo {note}
  {[Erratum: Phys.Rev.C 111, 029902 (2025)]},\ \Eprint
  {http://arxiv.org/abs/2101.12413} {arXiv:2101.12413 [nucl-ex]} \BibitemShut
  {NoStop}%
\bibitem [{\citenamefont {Aggarwal}\ \emph {et~al.}(2010)\citenamefont
  {Aggarwal} \emph {et~al.}}]{STAR:2010mib}%
  \BibitemOpen
  \bibfield  {author} {\bibinfo {author} {\bibfnamefont {M.~M.}\ \bibnamefont
  {Aggarwal}} \emph {et~al.} (\bibinfo {collaboration} {STAR}),\ }\href
  {\doibase 10.1103/PhysRevLett.105.022302} {\bibfield  {journal} {\bibinfo
  {journal} {Phys. Rev. Lett.}\ }\textbf {\bibinfo {volume} {105}},\ \bibinfo
  {pages} {022302} (\bibinfo {year} {2010})},\ \Eprint
  {http://arxiv.org/abs/1004.4959} {arXiv:1004.4959 [nucl-ex]} \BibitemShut
  {NoStop}%
\bibitem [{\citenamefont {Adamczyk}\ \emph {et~al.}(2014)\citenamefont
  {Adamczyk} \emph {et~al.}}]{STAR:2013gus}%
  \BibitemOpen
  \bibfield  {author} {\bibinfo {author} {\bibfnamefont {L.}~\bibnamefont
  {Adamczyk}} \emph {et~al.} (\bibinfo {collaboration} {STAR}),\ }\href
  {\doibase 10.1103/PhysRevLett.112.032302} {\bibfield  {journal} {\bibinfo
  {journal} {Phys. Rev. Lett.}\ }\textbf {\bibinfo {volume} {112}},\ \bibinfo
  {pages} {032302} (\bibinfo {year} {2014})},\ \Eprint
  {http://arxiv.org/abs/1309.5681} {arXiv:1309.5681 [nucl-ex]} \BibitemShut
  {NoStop}%
\bibitem [{\citenamefont {Adam}\ \emph {et~al.}(2021)\citenamefont {Adam} \emph
  {et~al.}}]{STAR:2020tga}%
  \BibitemOpen
  \bibfield  {author} {\bibinfo {author} {\bibfnamefont {J.}~\bibnamefont
  {Adam}} \emph {et~al.} (\bibinfo {collaboration} {STAR}),\ }\href {\doibase
  10.1103/PhysRevLett.126.092301} {\bibfield  {journal} {\bibinfo  {journal}
  {Phys. Rev. Lett.}\ }\textbf {\bibinfo {volume} {126}},\ \bibinfo {pages}
  {092301} (\bibinfo {year} {2021})},\ \Eprint
  {http://arxiv.org/abs/2001.02852} {arXiv:2001.02852 [nucl-ex]} \BibitemShut
  {NoStop}%
\bibitem [{\citenamefont {Abdallah}\ \emph
  {et~al.}(2021{\natexlab{b}})\citenamefont {Abdallah} \emph
  {et~al.}}]{STAR:2021rls}%
  \BibitemOpen
  \bibfield  {author} {\bibinfo {author} {\bibfnamefont {M.}~\bibnamefont
  {Abdallah}} \emph {et~al.} (\bibinfo {collaboration} {STAR}),\ }\href
  {\doibase 10.1103/PhysRevLett.127.262301} {\bibfield  {journal} {\bibinfo
  {journal} {Phys. Rev. Lett.}\ }\textbf {\bibinfo {volume} {127}},\ \bibinfo
  {pages} {262301} (\bibinfo {year} {2021}{\natexlab{b}})},\ \Eprint
  {http://arxiv.org/abs/2105.14698} {arXiv:2105.14698 [nucl-ex]} \BibitemShut
  {NoStop}%
\bibitem [{\citenamefont {Aboona}\ \emph {et~al.}(2023)\citenamefont {Aboona}
  \emph {et~al.}}]{STAR:2022vlo}%
  \BibitemOpen
  \bibfield  {author} {\bibinfo {author} {\bibfnamefont {B.}~\bibnamefont
  {Aboona}} \emph {et~al.} (\bibinfo {collaboration} {STAR}),\ }\href {\doibase
  10.1103/PhysRevLett.130.082301} {\bibfield  {journal} {\bibinfo  {journal}
  {Phys. Rev. Lett.}\ }\textbf {\bibinfo {volume} {130}},\ \bibinfo {pages}
  {082301} (\bibinfo {year} {2023})},\ \Eprint
  {http://arxiv.org/abs/2207.09837} {arXiv:2207.09837 [nucl-ex]} \BibitemShut
  {NoStop}%
\bibitem [{\citenamefont {Abdallah}\ \emph {et~al.}(2022)\citenamefont
  {Abdallah} \emph {et~al.}}]{STAR:2021fge}%
  \BibitemOpen
  \bibfield  {author} {\bibinfo {author} {\bibfnamefont {M.~S.}\ \bibnamefont
  {Abdallah}} \emph {et~al.} (\bibinfo {collaboration} {STAR}),\ }\href
  {\doibase 10.1103/PhysRevLett.128.202303} {\bibfield  {journal} {\bibinfo
  {journal} {Phys. Rev. Lett.}\ }\textbf {\bibinfo {volume} {128}},\ \bibinfo
  {pages} {202303} (\bibinfo {year} {2022})},\ \Eprint
  {http://arxiv.org/abs/2112.00240} {arXiv:2112.00240 [nucl-ex]} \BibitemShut
  {NoStop}%
\bibitem [{\citenamefont {Abdallah}\ \emph {et~al.}(2023)\citenamefont
  {Abdallah} \emph {et~al.}}]{STAR:2022etb}%
  \BibitemOpen
  \bibfield  {author} {\bibinfo {author} {\bibfnamefont {M.}~\bibnamefont
  {Abdallah}} \emph {et~al.} (\bibinfo {collaboration} {STAR}),\ }\href
  {\doibase 10.1103/PhysRevC.107.024908} {\bibfield  {journal} {\bibinfo
  {journal} {Phys. Rev. C}\ }\textbf {\bibinfo {volume} {107}},\ \bibinfo
  {pages} {024908} (\bibinfo {year} {2023})},\ \Eprint
  {http://arxiv.org/abs/2209.11940} {arXiv:2209.11940 [nucl-ex]} \BibitemShut
  {NoStop}%
\bibitem [{\citenamefont {SN0644}()}]{ref_iTPC_BESII}%
  \BibitemOpen
  \bibfield  {author} {\bibinfo {author} {\bibnamefont {SN0644}} (\bibinfo
  {collaboration} {STAR}),\ }\href
  {https://drupal.star.bnl.gov/STAR/starnotes/public/sn0644} {\bibinfo
  {journal} {Technical Design Report for the iTPC Upgrade}\ }\BibitemShut
  {NoStop}%
\bibitem [{\citenamefont {Bzdak}\ \emph {et~al.}(2017)\citenamefont {Bzdak},
  \citenamefont {Koch},\ and\ \citenamefont {Strodthoff}}]{Bzdak:2016sxg}%
  \BibitemOpen
\bibfield  {journal} {  }\bibfield  {author} {\bibinfo {author} {\bibfnamefont
  {A.}~\bibnamefont {Bzdak}}, \bibinfo {author} {\bibfnamefont
  {V.}~\bibnamefont {Koch}}, \ and\ \bibinfo {author} {\bibfnamefont
  {N.}~\bibnamefont {Strodthoff}},\ }\href {\doibase
  10.1103/PhysRevC.95.054906} {\bibfield  {journal} {\bibinfo  {journal} {Phys.
  Rev. C}\ }\textbf {\bibinfo {volume} {95}},\ \bibinfo {pages} {054906}
  (\bibinfo {year} {2017})},\ \Eprint {http://arxiv.org/abs/1607.07375}
  {arXiv:1607.07375 [nucl-th]} \BibitemShut {NoStop}%
\bibitem [{\citenamefont {Adler}\ \emph
  {et~al.}(2001{\natexlab{a}})\citenamefont {Adler}, \citenamefont {Denisov},
  \citenamefont {Garcia}, \citenamefont {Murray}, \citenamefont {Strobele},\
  and\ \citenamefont {White}}]{Adler:2000bd}%
  \BibitemOpen
  \bibfield  {author} {\bibinfo {author} {\bibfnamefont {C.}~\bibnamefont
  {Adler}}, \bibinfo {author} {\bibfnamefont {A.}~\bibnamefont {Denisov}},
  \bibinfo {author} {\bibfnamefont {E.}~\bibnamefont {Garcia}}, \bibinfo
  {author} {\bibfnamefont {M.~J.}\ \bibnamefont {Murray}}, \bibinfo {author}
  {\bibfnamefont {H.}~\bibnamefont {Strobele}}, \ and\ \bibinfo {author}
  {\bibfnamefont {S.~N.}\ \bibnamefont {White}},\ }\href {\doibase
  10.1016/S0168-9002(01)00627-1} {\bibfield  {journal} {\bibinfo  {journal}
  {Nucl. Instrum. Meth. A}\ }\textbf {\bibinfo {volume} {470}},\ \bibinfo
  {pages} {488} (\bibinfo {year} {2001}{\natexlab{a}})},\ \Eprint
  {http://arxiv.org/abs/nucl-ex/0008005} {arXiv:nucl-ex/0008005} \BibitemShut
  {NoStop}%
\bibitem [{\citenamefont {Llope}\ \emph {et~al.}(2004)\citenamefont {Llope}
  \emph {et~al.}}]{Llope:2003ti}%
  \BibitemOpen
  \bibfield  {author} {\bibinfo {author} {\bibfnamefont {W.~J.}\ \bibnamefont
  {Llope}} \emph {et~al.},\ }\href {\doibase 10.1016/j.nima.2003.11.414}
  {\bibfield  {journal} {\bibinfo  {journal} {Nucl. Instrum. Meth. A}\ }\textbf
  {\bibinfo {volume} {522}},\ \bibinfo {pages} {252} (\bibinfo {year}
  {2004})},\ \Eprint {http://arxiv.org/abs/nucl-ex/0308022}
  {arXiv:nucl-ex/0308022} \BibitemShut {NoStop}%
\bibitem [{\citenamefont {Ackermann}\ \emph {et~al.}(2003)\citenamefont
  {Ackermann} \emph {et~al.}}]{STAR:2002eio}%
  \BibitemOpen
  \bibfield  {author} {\bibinfo {author} {\bibfnamefont {K.~H.}\ \bibnamefont
  {Ackermann}} \emph {et~al.} (\bibinfo {collaboration} {STAR}),\ }\href
  {\doibase 10.1016/S0168-9002(02)01960-5} {\bibfield  {journal} {\bibinfo
  {journal} {Nucl. Instrum. Meth. A}\ }\textbf {\bibinfo {volume} {499}},\
  \bibinfo {pages} {624} (\bibinfo {year} {2003})}\BibitemShut {NoStop}%
\bibitem [{\citenamefont {Adler}\ \emph
  {et~al.}(2001{\natexlab{b}})\citenamefont {Adler} \emph
  {et~al.}}]{STAR:2001rbj}%
  \BibitemOpen
  \bibfield  {author} {\bibinfo {author} {\bibfnamefont {C.}~\bibnamefont
  {Adler}} \emph {et~al.} (\bibinfo {collaboration} {STAR}),\ }\href {\doibase
  10.1103/PhysRevLett.86.4778} {\bibfield  {journal} {\bibinfo  {journal}
  {Phys. Rev. Lett.}\ }\textbf {\bibinfo {volume} {86}},\ \bibinfo {pages}
  {4778} (\bibinfo {year} {2001}{\natexlab{b}})},\ \bibinfo {note} {[Erratum:
  Phys.Rev.Lett. 90, 119903 (2003)]},\ \Eprint
  {http://arxiv.org/abs/nucl-ex/0104022} {arXiv:nucl-ex/0104022} \BibitemShut
  {NoStop}%
\bibitem [{\citenamefont {Vovchenko}\ \emph {et~al.}(2022)\citenamefont
  {Vovchenko}, \citenamefont {Koch},\ and\ \citenamefont
  {Shen}}]{Vovchenko:2021kxx}%
  \BibitemOpen
  \bibfield  {author} {\bibinfo {author} {\bibfnamefont {V.}~\bibnamefont
  {Vovchenko}}, \bibinfo {author} {\bibfnamefont {V.}~\bibnamefont {Koch}}, \
  and\ \bibinfo {author} {\bibfnamefont {C.}~\bibnamefont {Shen}},\ }\href
  {\doibase 10.1103/PhysRevC.105.014904} {\bibfield  {journal} {\bibinfo
  {journal} {Phys. Rev. C}\ }\textbf {\bibinfo {volume} {105}},\ \bibinfo
  {pages} {014904} (\bibinfo {year} {2022})},\ \Eprint
  {http://arxiv.org/abs/2107.00163} {arXiv:2107.00163 [hep-ph]} \BibitemShut
  {NoStop}%
\bibitem [{\citenamefont {Zhang}\ \emph {et~al.}(2020)\citenamefont {Zhang},
  \citenamefont {He}, \citenamefont {Liu}, \citenamefont {Yang},\ and\
  \citenamefont {Luo}}]{Zhang:2019lqz}%
  \BibitemOpen
  \bibfield  {author} {\bibinfo {author} {\bibfnamefont {Y.}~\bibnamefont
  {Zhang}}, \bibinfo {author} {\bibfnamefont {S.}~\bibnamefont {He}}, \bibinfo
  {author} {\bibfnamefont {H.}~\bibnamefont {Liu}}, \bibinfo {author}
  {\bibfnamefont {Z.}~\bibnamefont {Yang}}, \ and\ \bibinfo {author}
  {\bibfnamefont {X.}~\bibnamefont {Luo}},\ }\href {\doibase
  10.1103/PhysRevC.101.034909} {\bibfield  {journal} {\bibinfo  {journal}
  {Phys. Rev. C}\ }\textbf {\bibinfo {volume} {101}},\ \bibinfo {pages}
  {034909} (\bibinfo {year} {2020})},\ \Eprint
  {http://arxiv.org/abs/1905.01095} {arXiv:1905.01095 [nucl-ex]} \BibitemShut
  {NoStop}%
\bibitem [{\citenamefont {Chatterjee}\ \emph {et~al.}(2020)\citenamefont
  {Chatterjee}, \citenamefont {Zhang}, \citenamefont {Zeng}, \citenamefont
  {Sahoo},\ and\ \citenamefont {Luo}}]{Chatterjee:2019fey}%
  \BibitemOpen
  \bibfield  {author} {\bibinfo {author} {\bibfnamefont {A.}~\bibnamefont
  {Chatterjee}}, \bibinfo {author} {\bibfnamefont {Y.}~\bibnamefont {Zhang}},
  \bibinfo {author} {\bibfnamefont {J.}~\bibnamefont {Zeng}}, \bibinfo {author}
  {\bibfnamefont {N.~R.}\ \bibnamefont {Sahoo}}, \ and\ \bibinfo {author}
  {\bibfnamefont {X.}~\bibnamefont {Luo}},\ }\href {\doibase
  10.1103/PhysRevC.101.034902} {\bibfield  {journal} {\bibinfo  {journal}
  {Phys. Rev. C}\ }\textbf {\bibinfo {volume} {101}},\ \bibinfo {pages}
  {034902} (\bibinfo {year} {2020})},\ \Eprint
  {http://arxiv.org/abs/1910.08004} {arXiv:1910.08004 [nucl-ex]} \BibitemShut
  {NoStop}%
\bibitem [{\citenamefont {Luo}\ and\ \citenamefont {Xu}(2017)}]{Luo:2017faz}%
  \BibitemOpen
  \bibfield  {author} {\bibinfo {author} {\bibfnamefont {X.}~\bibnamefont
  {Luo}}\ and\ \bibinfo {author} {\bibfnamefont {N.}~\bibnamefont {Xu}},\
  }\href {\doibase 10.1007/s41365-017-0257-0} {\bibfield  {journal} {\bibinfo
  {journal} {Nucl. Sci. Tech.}\ }\textbf {\bibinfo {volume} {28}},\ \bibinfo
  {pages} {112} (\bibinfo {year} {2017})},\ \Eprint
  {http://arxiv.org/abs/1701.02105} {arXiv:1701.02105 [nucl-ex]} \BibitemShut
  {NoStop}%
\bibitem [{\citenamefont {Luo}\ \emph {et~al.}(2013)\citenamefont {Luo},
  \citenamefont {Xu}, \citenamefont {Mohanty},\ and\ \citenamefont
  {Xu}}]{Luo:2013bmi}%
  \BibitemOpen
  \bibfield  {author} {\bibinfo {author} {\bibfnamefont {X.}~\bibnamefont
  {Luo}}, \bibinfo {author} {\bibfnamefont {J.}~\bibnamefont {Xu}}, \bibinfo
  {author} {\bibfnamefont {B.}~\bibnamefont {Mohanty}}, \ and\ \bibinfo
  {author} {\bibfnamefont {N.}~\bibnamefont {Xu}},\ }\href {\doibase
  10.1088/0954-3899/40/10/105104} {\bibfield  {journal} {\bibinfo  {journal}
  {J. Phys. G}\ }\textbf {\bibinfo {volume} {40}},\ \bibinfo {pages} {105104}
  (\bibinfo {year} {2013})},\ \Eprint {http://arxiv.org/abs/1302.2332}
  {arXiv:1302.2332 [nucl-ex]} \BibitemShut {NoStop}%
\bibitem [{ref()}]{ref_suppl_material}%
  \BibitemOpen
  \href@noop {} {\bibinfo  {journal} {{See Supplemental Material at [LINK] for
  CBWC, effect of centrality resolution on cumulant ratios, comparison of
  net-proton $C_4/C_2$ data from BES-II vs. BES-I and with various baselines}}\
  }\BibitemShut {NoStop}%
\bibitem [{\citenamefont {Bzdak}\ and\ \citenamefont
  {Koch}(2012)}]{Bzdak:2012ab}%
  \BibitemOpen
\bibfield  {journal} {  }\bibfield  {author} {\bibinfo {author} {\bibfnamefont
  {A.}~\bibnamefont {Bzdak}}\ and\ \bibinfo {author} {\bibfnamefont
  {V.}~\bibnamefont {Koch}},\ }\href {\doibase 10.1103/PhysRevC.86.044904}
  {\bibfield  {journal} {\bibinfo  {journal} {Phys. Rev. C}\ }\textbf {\bibinfo
  {volume} {86}},\ \bibinfo {pages} {044904} (\bibinfo {year} {2012})},\
  \Eprint {http://arxiv.org/abs/1206.4286} {arXiv:1206.4286 [nucl-th]}
  \BibitemShut {NoStop}%
\bibitem [{\citenamefont {Kitazawa}\ and\ \citenamefont
  {Asakawa}(2012)}]{Kitazawa:2012at}%
  \BibitemOpen
  \bibfield  {author} {\bibinfo {author} {\bibfnamefont {M.}~\bibnamefont
  {Kitazawa}}\ and\ \bibinfo {author} {\bibfnamefont {M.}~\bibnamefont
  {Asakawa}},\ }\href {\doibase 10.1103/PhysRevC.86.024904} {\bibfield
  {journal} {\bibinfo  {journal} {Phys. Rev. C}\ }\textbf {\bibinfo {volume}
  {86}},\ \bibinfo {pages} {024904} (\bibinfo {year} {2012})},\ \bibinfo {note}
  {[Erratum: Phys.Rev.C 86, 069902 (2012)]},\ \Eprint
  {http://arxiv.org/abs/1205.3292} {arXiv:1205.3292 [nucl-th]} \BibitemShut
  {NoStop}%
\bibitem [{\citenamefont {Luo}(2015)}]{Luo:2014rea}%
  \BibitemOpen
  \bibfield  {author} {\bibinfo {author} {\bibfnamefont {X.}~\bibnamefont
  {Luo}},\ }\href {\doibase 10.1103/PhysRevC.94.059901} {\bibfield  {journal}
  {\bibinfo  {journal} {Phys. Rev. C}\ }\textbf {\bibinfo {volume} {91}},\
  \bibinfo {pages} {034907} (\bibinfo {year} {2015})},\ \bibinfo {note}
  {[Erratum: Phys.Rev.C 94, 059901 (2016)]},\ \Eprint
  {http://arxiv.org/abs/1410.3914} {arXiv:1410.3914 [physics.data-an]}
  \BibitemShut {NoStop}%
\bibitem [{\citenamefont {Nonaka}\ \emph {et~al.}(2017)\citenamefont {Nonaka},
  \citenamefont {Kitazawa},\ and\ \citenamefont {Esumi}}]{Nonaka:2017kko}%
  \BibitemOpen
  \bibfield  {author} {\bibinfo {author} {\bibfnamefont {T.}~\bibnamefont
  {Nonaka}}, \bibinfo {author} {\bibfnamefont {M.}~\bibnamefont {Kitazawa}}, \
  and\ \bibinfo {author} {\bibfnamefont {S.}~\bibnamefont {Esumi}},\ }\href
  {\doibase 10.1103/PhysRevC.95.064912} {\bibfield  {journal} {\bibinfo
  {journal} {Phys. Rev. C}\ }\textbf {\bibinfo {volume} {95}},\ \bibinfo
  {pages} {064912} (\bibinfo {year} {2017})},\ \bibinfo {note} {[Erratum:
  Phys.Rev.C 103, 029901 (2021)]},\ \Eprint {http://arxiv.org/abs/1702.07106}
  {arXiv:1702.07106 [physics.data-an]} \BibitemShut {NoStop}%
\bibitem [{\citenamefont {Luo}\ and\ \citenamefont
  {Nonaka}(2019)}]{Luo:2018ofd}%
  \BibitemOpen
  \bibfield  {author} {\bibinfo {author} {\bibfnamefont {X.}~\bibnamefont
  {Luo}}\ and\ \bibinfo {author} {\bibfnamefont {T.}~\bibnamefont {Nonaka}},\
  }\href {\doibase 10.1103/PhysRevC.99.044917} {\bibfield  {journal} {\bibinfo
  {journal} {Phys. Rev. C}\ }\textbf {\bibinfo {volume} {99}},\ \bibinfo
  {pages} {044917} (\bibinfo {year} {2019})},\ \Eprint
  {http://arxiv.org/abs/1812.10303} {arXiv:1812.10303 [physics.data-an]}
  \BibitemShut {NoStop}%
\bibitem [{\citenamefont {Barlow}(2002)}]{Barlow:2002yb}%
  \BibitemOpen
  \bibfield  {author} {\bibinfo {author} {\bibfnamefont {R.}~\bibnamefont
  {Barlow}},\ }in\ \href@noop {} {\emph {\bibinfo {booktitle} {{Conference on
  Advanced Statistical Techniques in Particle Physics}}}}\ (\bibinfo {year}
  {2002})\ pp.\ \bibinfo {pages} {134--144},\ \Eprint
  {http://arxiv.org/abs/hep-ex/0207026} {arXiv:hep-ex/0207026} \BibitemShut
  {NoStop}%
\bibitem [{\citenamefont {Braun-Munzinger}\ \emph {et~al.}(2021)\citenamefont
  {Braun-Munzinger}, \citenamefont {Friman}, \citenamefont {Redlich},
  \citenamefont {Rustamov},\ and\ \citenamefont
  {Stachel}}]{Braun-Munzinger:2020jbk}%
  \BibitemOpen
  \bibfield  {author} {\bibinfo {author} {\bibfnamefont {P.}~\bibnamefont
  {Braun-Munzinger}}, \bibinfo {author} {\bibfnamefont {B.}~\bibnamefont
  {Friman}}, \bibinfo {author} {\bibfnamefont {K.}~\bibnamefont {Redlich}},
  \bibinfo {author} {\bibfnamefont {A.}~\bibnamefont {Rustamov}}, \ and\
  \bibinfo {author} {\bibfnamefont {J.}~\bibnamefont {Stachel}},\ }\href
  {\doibase 10.1016/j.nuclphysa.2021.122141} {\bibfield  {journal} {\bibinfo
  {journal} {Nucl. Phys. A}\ }\textbf {\bibinfo {volume} {1008}},\ \bibinfo
  {pages} {122141} (\bibinfo {year} {2021})},\ \Eprint
  {http://arxiv.org/abs/2007.02463} {arXiv:2007.02463 [nucl-th]} \BibitemShut
  {NoStop}%
\bibitem [{\citenamefont {Bass}\ \emph {et~al.}(1998)\citenamefont {Bass} \emph
  {et~al.}}]{Bass:1998ca}%
  \BibitemOpen
  \bibfield  {author} {\bibinfo {author} {\bibfnamefont {S.~A.}\ \bibnamefont
  {Bass}} \emph {et~al.},\ }\href {\doibase 10.1016/S0146-6410(98)00058-1}
  {\bibfield  {journal} {\bibinfo  {journal} {Prog. Part. Nucl. Phys.}\
  }\textbf {\bibinfo {volume} {41}},\ \bibinfo {pages} {255} (\bibinfo {year}
  {1998})},\ \Eprint {http://arxiv.org/abs/nucl-th/9803035}
  {arXiv:nucl-th/9803035} \BibitemShut {NoStop}%
\bibitem [{\citenamefont {Bleicher}\ \emph {et~al.}(1999)\citenamefont
  {Bleicher} \emph {et~al.}}]{Bleicher:1999xi}%
  \BibitemOpen
  \bibfield  {author} {\bibinfo {author} {\bibfnamefont {M.}~\bibnamefont
  {Bleicher}} \emph {et~al.},\ }\href {\doibase 10.1088/0954-3899/25/9/308}
  {\bibfield  {journal} {\bibinfo  {journal} {J. Phys. G}\ }\textbf {\bibinfo
  {volume} {25}},\ \bibinfo {pages} {1859} (\bibinfo {year} {1999})},\ \Eprint
  {http://arxiv.org/abs/hep-ph/9909407} {arXiv:hep-ph/9909407} \BibitemShut
  {NoStop}%
\bibitem [{\citenamefont {Bazavov}\ \emph {et~al.}(2020)\citenamefont {Bazavov}
  \emph {et~al.}}]{Bazavov:2020bjn}%
  \BibitemOpen
  \bibfield  {author} {\bibinfo {author} {\bibfnamefont {A.}~\bibnamefont
  {Bazavov}} \emph {et~al.},\ }\href {\doibase 10.1103/PhysRevD.101.074502}
  {\bibfield  {journal} {\bibinfo  {journal} {Phys. Rev. D}\ }\textbf {\bibinfo
  {volume} {101}},\ \bibinfo {pages} {074502} (\bibinfo {year} {2020})},\
  \Eprint {http://arxiv.org/abs/2001.08530} {arXiv:2001.08530 [hep-lat]}
  \BibitemShut {NoStop}%
\bibitem [{\citenamefont {Bollweg}\ \emph {et~al.}(2024)\citenamefont
  {Bollweg}, \citenamefont {Ding}, \citenamefont {Goswami}, \citenamefont
  {Karsch}, \citenamefont {Mukherjee}, \citenamefont {Petreczky},\ and\
  \citenamefont {Schmidt}}]{Bollweg:2024epj}%
  \BibitemOpen
  \bibfield  {author} {\bibinfo {author} {\bibfnamefont {D.}~\bibnamefont
  {Bollweg}}, \bibinfo {author} {\bibfnamefont {H.~T.}\ \bibnamefont {Ding}},
  \bibinfo {author} {\bibfnamefont {J.}~\bibnamefont {Goswami}}, \bibinfo
  {author} {\bibfnamefont {F.}~\bibnamefont {Karsch}}, \bibinfo {author}
  {\bibfnamefont {S.}~\bibnamefont {Mukherjee}}, \bibinfo {author}
  {\bibfnamefont {P.}~\bibnamefont {Petreczky}}, \ and\ \bibinfo {author}
  {\bibfnamefont {C.}~\bibnamefont {Schmidt}},\ }\href@noop {} {\  (\bibinfo
  {year} {2024})},\ \Eprint {http://arxiv.org/abs/2407.09335} {arXiv:2407.09335
  [hep-lat]} \BibitemShut {NoStop}%
\bibitem [{\citenamefont {Adamczyk}\ \emph {et~al.}(2017)\citenamefont
  {Adamczyk} \emph {et~al.}}]{STAR:2017sal}%
  \BibitemOpen
  \bibfield  {author} {\bibinfo {author} {\bibfnamefont {L.}~\bibnamefont
  {Adamczyk}} \emph {et~al.} (\bibinfo {collaboration} {STAR}),\ }\href
  {\doibase 10.1103/PhysRevC.96.044904} {\bibfield  {journal} {\bibinfo
  {journal} {Phys. Rev. C}\ }\textbf {\bibinfo {volume} {96}},\ \bibinfo
  {pages} {044904} (\bibinfo {year} {2017})},\ \Eprint
  {http://arxiv.org/abs/1701.07065} {arXiv:1701.07065 [nucl-ex]} \BibitemShut
  {NoStop}%
\bibitem [{\citenamefont {An}\ \emph {et~al.}(2022)\citenamefont {An} \emph
  {et~al.}}]{An:2021wof}%
  \BibitemOpen
  \bibfield  {author} {\bibinfo {author} {\bibfnamefont {X.}~\bibnamefont {An}}
  \emph {et~al.},\ }\href {\doibase 10.1016/j.nuclphysa.2021.122343} {\bibfield
   {journal} {\bibinfo  {journal} {Nucl. Phys. A}\ }\textbf {\bibinfo {volume}
  {1017}},\ \bibinfo {pages} {122343} (\bibinfo {year} {2022})},\ \Eprint
  {http://arxiv.org/abs/2108.13867} {arXiv:2108.13867 [nucl-th]} \BibitemShut
  {NoStop}%
\end{thebibliography}%

\clearpage
\section{Supplemental Material}

\subsection{Effect of CBWC correction on net-proton cumulant ratios}
In this letter, an event-weighted averaging of measurements over multiplicity bins, called the centrality-bin-width correction (CBWC) is performed when reporting results for a given centrality class, for example, 0-5\%. The underlying idea is that the mean ($C_1$) of a net-proton distribution strongly varies across RefMult3X multiplicity bins within a centrality class. Since $C_1$ appears in all higher order cumulants and their ratios, proper averaging should ensure that this multiplicity dependence of the mean is taken care of. The CBWC ensures this by calculating cumulants in each multiplicity bin and performing a weighted average across the multiplicity bins where the number of events are taken as the weights. Without such an averaging in place, the cumulants and their ratios in a given centrality fall off from the multiplicity dependence trend. They also become dependent on the width of the centrality class. Demonstration of such a behavior for net-proton cumulant ratios in Au+Au collisions at $\sqrt{s_{NN}}=19.6$ GeV from BES-II can be seen in Fig.~\ref{labl_figS1}. 
\setcounter{figure}{0}
\begin{figure}[!htbp]
	\centering
	\includegraphics[scale=0.4]{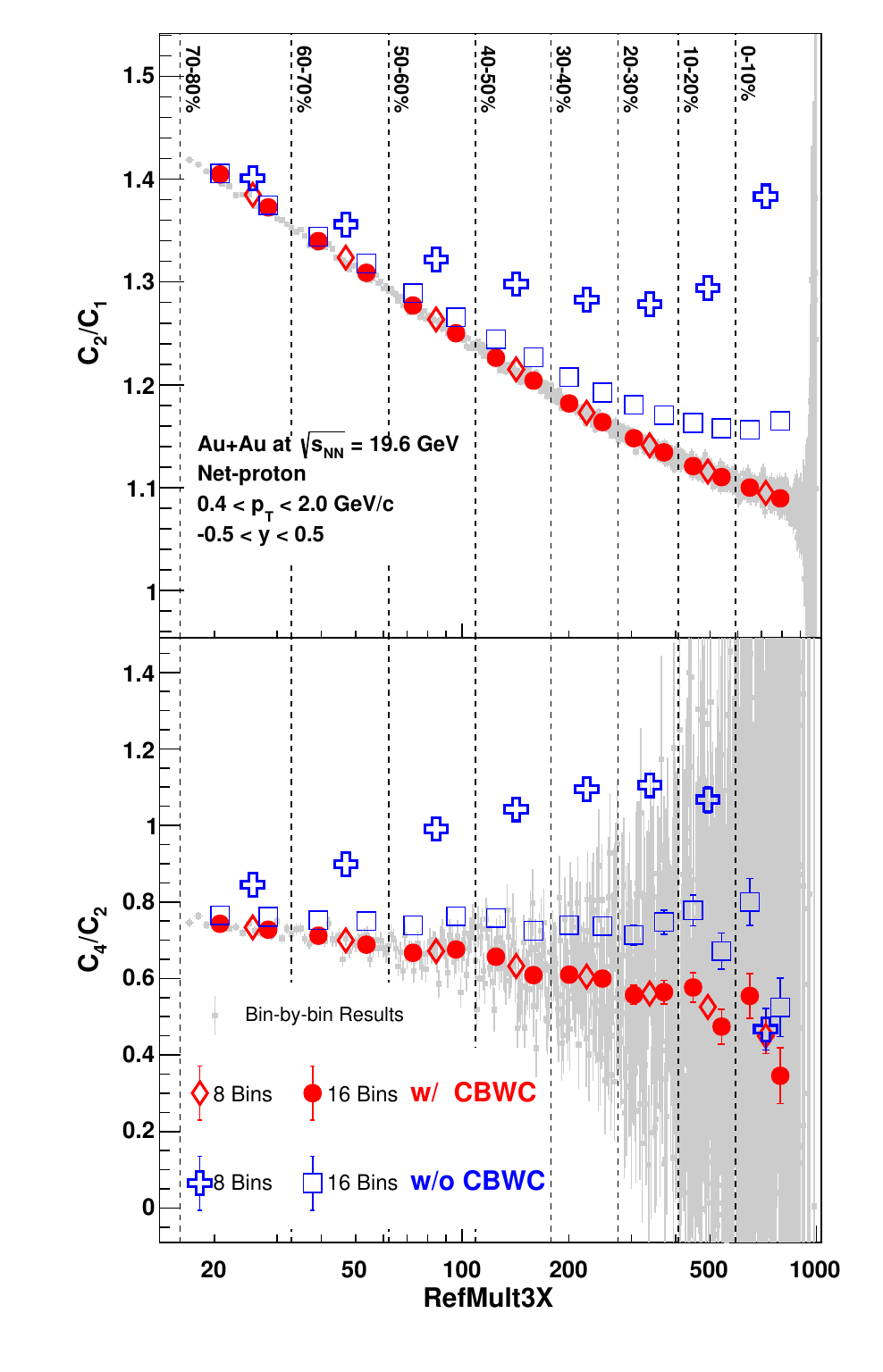}
	\caption{Net-proton cumulant ratios $C_2/C_1$ and $C_4/C_2$ as a function of RefMult3X. The grey band represents their values including $\pm$ their statistical uncertainties at each unit multiplicity bin of RefMult3X. The results for 8 and 16 collision centrality classes from 0-5\% to 70-80\% represented by vertical dashed lines are shown. These results with and without the centrality-bin-width correction are shown as red and blue markers, respectively. The uncertainties are only statistical.   }
	\label{labl_figS1}
\end{figure}
Results in centrality classes with 5\% steps, such as 0-5\% and 5-10\%, and those with 10\% steps, such as 0-10\% and 10-20\% being of different centrality widths, show distinctly different centrality dependence. With the CBWC averaging done, independent of the centrality width, a smooth dependence of the measurements consistent with the trend from multiplicity dependence is observed.

\subsection{Effect of centrality resolution on cumulant ratios}
\begin{figure*}[!htbp]
	\centering
	\includegraphics[width=\linewidth]{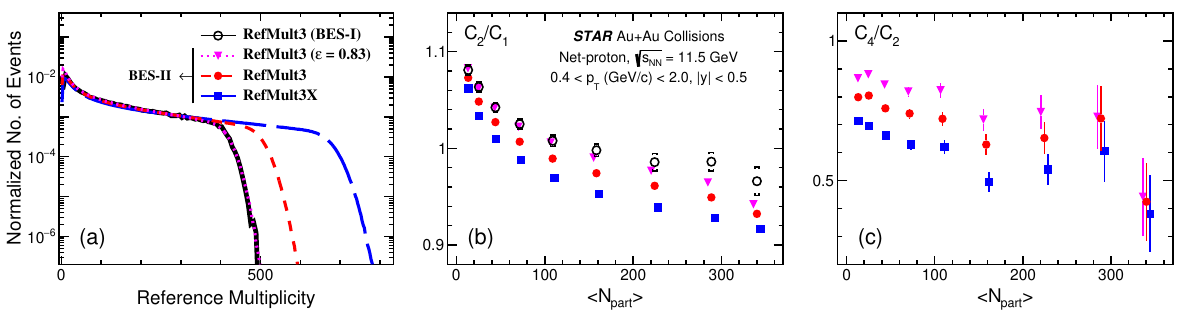}
	\caption{Panel(a): Distributions of reference charged particle multiplicities: BES-I RefMult3 (black open circles), BES-II RefMult3 (red solid circles) and RefMult3X (blue solid squares). A new multiplicity (magenta solid inverted triangles) is defined by applying a binomial efficiency of 83\% to each charged particle counted in BES-II RefMult3 to mimic the centrality resolution of BES-I RefMul3. Panels (b) and (c): Net-proton cumulant ratios $C_2/C_1$ and $C_4/C_2$ as a function of collision centrality from various multiplicities. A clear centrality resolution ordering is observed: a better centrality resolution results in lower cumulant ratios, while the $C_4/C_2$ from central events shows weak centrality resolution dependence. Uncertainties are statistical (except for BES-I where systemtic uncertainties are also presented as brackets). }
	\label{labl_figS2}
\end{figure*}
In the main text, two sets of results on net-proton cumulants from BES-II were discussed based on two centrality definitions: RefMult3 and RefMult3X where the difference was pseudorapidity acceptance, i.e, $|\eta|<1.0$ and $|\eta|<1.6$, respectively. In addition, a comparison was also made with BES-I results which used the RefMult3 definition for centrality. Figure~\ref{labl_figS2} panel (a) shows the three cases of centrality definition at $\sqrt{s_{NN}}=11.5$ GeV along with results on net-proton cumulant ratios $C_2/C_1$ and $C_4/C_2$. Note that RefMult3 from BES-I is smaller than RefMult3 from BES-II due to the iTPC upgrade. A clear ordering is seen in cumulant ratios: higher centrality resolution (resulting from larger multiplicity), leads to lower net-proton cumulant ratios except for $C_4/C_2$ in 0-5\% centrality where results were consistent within uncertainties irrespective of difference in resolution. An additional set of results is shown (RefMult3E) where random sampling with an efficiency factor of 83\% is introduced in the counting of RefMult3 in BES-II to mimic the RefMult3 distribution at BES-I. This is added to have a similar centrality resolution in BES-II data as that of BES-I. Comparison with results from RefMult3E suggests that once centrality resolutions are kept similar, the cumulant ratios show good consistency with those from BES-I.

\begin{figure*}[!htbp]
	\centering
	\includegraphics[width=\linewidth]{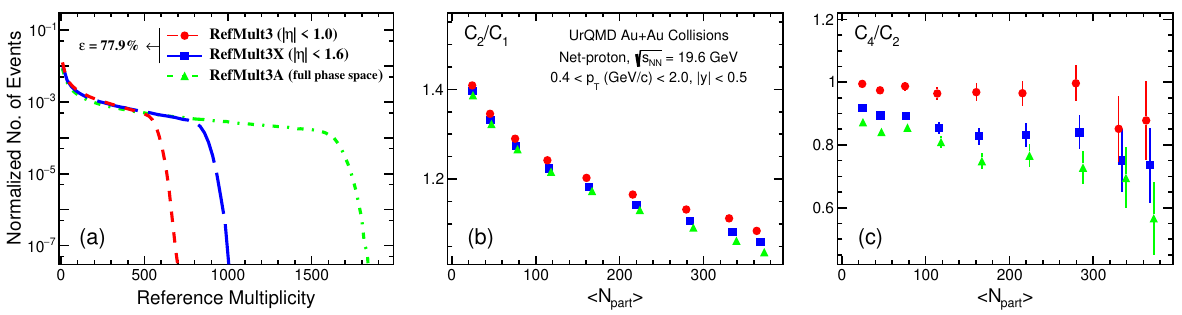}
	\caption{Panel(a): Distributions of reference charged particle multiplicities in UrQMD at $\sqrt{s_{NN}}=19.6$ GeV: RefMult3 (red filled circles), RefMult3X (blue filled squares). The two distributions are scaled to mimic the distribution in real data by applying an efficiency of 77.9\%. RefMult3A, which is the multiplicity distribution in the full phase space, is shown as green triangles. Panels (b) and (c): Net-proton cumulant ratios $C_2/C_1$ and $C_4/C_2$ as a function of collision centrality for the three centrality definitions. Errors on calculations are only statistical. }
	\label{labl_figS3}
\end{figure*}
Higher centrality resolution was observed to cause lower cumulant ratios. A UrQMD model study to test if the decreasing trend in cumulant ratios saturates with increasing centrality resolution is shown in Fig.~\ref{labl_figS3}. The case corresponding to the highest centrality resolution is when all charged particles in the phase space excluding protons are considered for centrality (RefMult3A). A saturation in the downward trend of ratios is indeed observed with increasing multiplicity. For example, while results from RefMult3A and the ones using RefMult3 show a deviation within 13.4\% (1.8\%) for $C_4/C_2$ ($C_2/C_1$) in 50-60\% centrality, those obtained with RefMult3X are much closer to the former and are only within 4.2\% (0.6\%) of each other for $C_4/C_2$ ($C_2/C_1$) for the same centrality.

\subsection{Comparison of net-proton $C_4/C_2$ measurements from BES-II vs BES-I}
\begin{figure}[!htbp]
	\centering
	\includegraphics[scale=0.55]{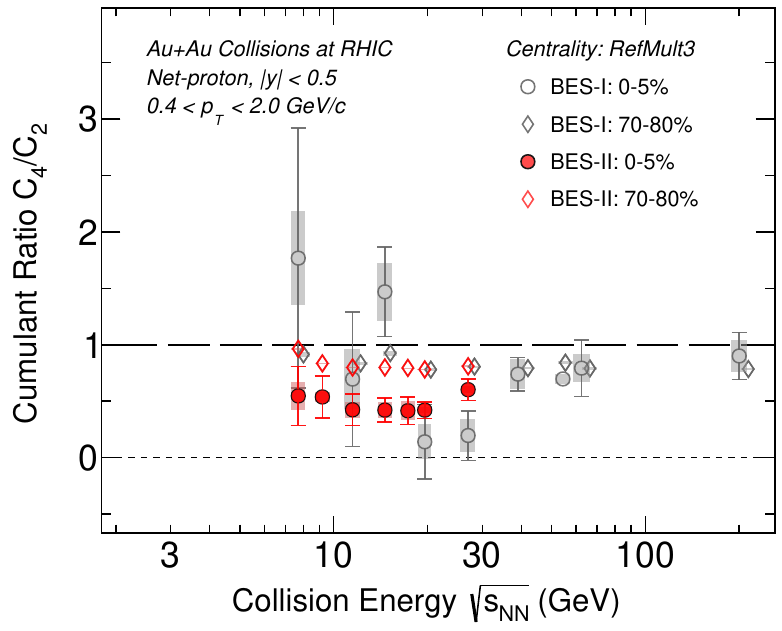}
	\caption{Net-proton cumulant ratios $C_4/C_2$ as a function of collision energy in Au+Au collisions at RHIC. The latest results from BES-II (red markers) with RefMult3 centrality definition are compared with those from BES-I (grey markers). The 0-5\% central collision results are shown as open circles and peripheral 70-80\% data as open diamonds. The peripheral datapoints from BES-I are slightly shifted in x-axis for clarity of presentation. The bars and bands on the data points reflect statistical and systematic uncertainties, respectively.}
	\label{labl_figS4}
\end{figure}

\setcounter{table}{0}
	\begin{table}[!htb]
		\caption{Significance of deviation between BES-II and BES-I net-proton $C_4/C_2$ data for 0-5\% and 70-80\% centrality at collision energies common to both programs. }
		\centering   
		\begin{tabular}{|c|c|c|}
			\hline	
			$\sqrt{s_{NN}}$ (GeV) &  0-5\% &  70-80\%  \\
			\hline 
			7.7   &  $1.0\sigma$ &  $0.9\sigma$  \\
			\hline 
			11.5   &  $0.4\sigma$ &  $1.3\sigma$  \\
			\hline 
			14.6   &  $2.2\sigma$ &  $2.5\sigma$  \\
			\hline 
			19.6   &  $0.8\sigma$ &  $0.1\sigma$  \\
			\hline 
			27   &  $1.4\sigma$ &  $0.2\sigma$  \\
			\hline 
		\end{tabular}
		\label{labl_besIvsII}
	\end{table}

Figure~\ref{labl_figS4} shows a comparison of net-proton cumulant ratio $C_4/C_2$ from BES-II vs BES-I in the most central (0-5\%) and peripheral (70-80\%) collisions. The BES-II results are found to be mostly consistent with BES-I within $\pm 1\sigma$ of total uncertainties. Table~\ref{labl_besIvsII} gives this deviation at each collision energy. A $\chi^2$ test from the comparison of BES-II and BES-I data from $\sqrt{s_{NN}}$ = 7.7 to 27 GeV yields a $p-$value of 0.136 (0.122) for 0-5\% ( 70-80\%) centrality, suggesting there are no significant differences between them.

\subsection{Difference between net-proton $C_4/C_2$ (0-5\%) and non-critical baselines}
\begin{figure}[!htbp]
	\centering
	\includegraphics[scale=0.4]{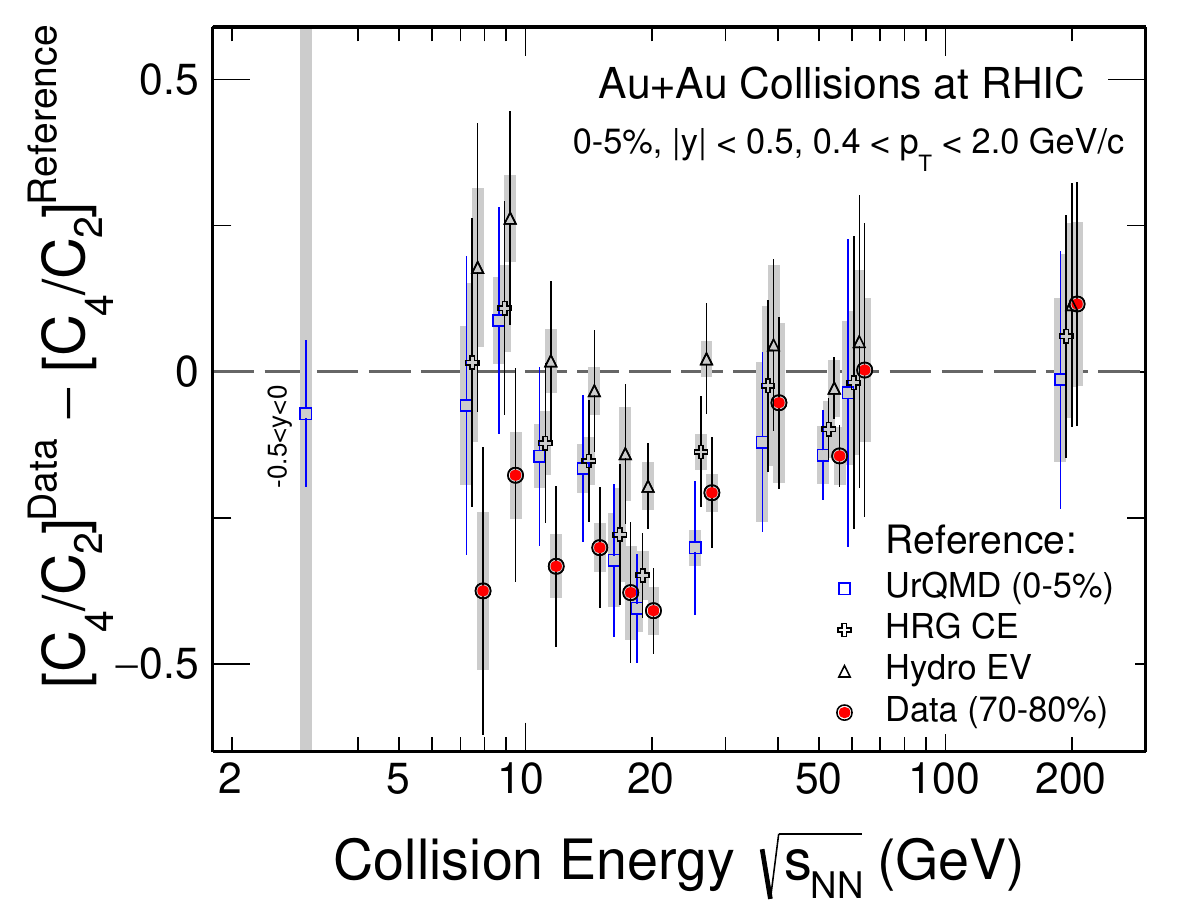}
	\caption{Difference between net-proton $C_4/C_2$ in 0-5\% central collisions and non-critical baselines (UrQMD, HRG CE, and hydro EV) as well as peripheral 70-80\% data as a function of collision energy. The values are marginally shifted along the x-axis for different choices of baselines for clarity of presentation. The data at $\sqrt{s_{NN}}=3$ GeV is with protons measured in rapidity acceptance of $-0.5<y<0$ while all other collision energies are measured with $|y|<0.5$. The bars and bands on the data points reflect statistical and systematic uncertainties, respectively.}
	\label{labl_figS5}
\end{figure}
Figure~\ref{labl_figS5} shows difference between net-proton $C_4/C_2$ in 0-5\% central collisions and non-critical baselines, as well as peripheral 70-80\% data. A minimum in collision energy dependence is observed at around $\sqrt{s_{NN}}=19.6$ GeV with deviation of $\sim2-5\sigma$ from zero.

\end{document}